\documentclass[aps,prd,eqsecnum,preprint]{revtex4}
\usepackage[dvipdfmx]{color,graphicx}
\usepackage{amsfonts}
\usepackage{amsmath}
\usepackage{amssymb}
\usepackage{bm}

\newcommand{\bH}{\ensuremath{\bar{H}}}
\newcommand{\brho}{\ensuremath{\bar{\rho}}}
\newcommand{\bk}{\ensuremath{{\bar{k}}}}

\newcommand{\hH}{\ensuremath{\hat{H}}}
\newcommand{\ha}{\ensuremath{\hat{a}}}
\newcommand{\hrho}{\ensuremath{\hat{\rho}}}
\newcommand{\hpsi}{\ensuremath{\hat{\psi}}}
\newcommand{\hD}{\ensuremath{\hat{D}}}

\newcommand{\fDelta}{\ensuremath{\Delta_{\rm f}}}

\begin{document}

{\baselineskip0pt
\leftline{\large\baselineskip16pt\sl\vbox to0pt{\hbox{\it Department of
Mathematics and Physics}
               \hbox{\it Osaka City  University}\vss}}
\rightline{\large\baselineskip16pt\rm\vbox to20pt{\hbox{OCU-PHYS-424}
            \hbox{AP-GR-124}
\vss}}%
}

\vskip1cm
\title{
Systematic error due to isotropic inhomogeneities
}

\author{
$^{1}$Hiroyuki Negishi\footnote{Electronic address:negishi@sci.osaka-cu.ac.jp}, 
$^{1}$Ken-ichi Nakao\footnote{Electronic address:knakao@sci.osaka-cu.ac.jp},
$^{2}$Chul-Moon Yoo\footnote{Electronic address:yoo@gravity.phys.nagoya-u.ac.jp }
and 
$^{1}$Ryusuke Nishikawa\footnote{Electronic address:ryusuke@sci.osaka-cu.ac.jp}
}

\affiliation{
$^{1}$Department of Mathematics and Physics,
Graduate School of Science, Osaka City University, Sumiyoshi-ku, 
Osaka City 558-8585, Japan\\
$^{2}$Division of Particle and Astrophysical Science,
Graduate School of Science, Nagoya University, Furo-cho, Chikusa-ku, Nagoya 464-8602, Japan
}
\date{\today}

\begin{abstract}
Usually the effects of isotropic inhomogeneities are not seriously taken into account 
in the determination of the cosmological parameters because of Copernican principle whose 
statement is that we do not live in the privileged domain in the universe.   
But Copernican principle has not been observationally confirmed yet in 
sufficient accuracy, and there is the possibility that there are non-negligible large-scale isotropic 
inhomogeneities in our universe. In this paper, we study the effects of the isotropic inhomogeneities 
on the determination of the cosmological parameters and 
show the probability that 
non-Copernican isotropic inhomogeneities mislead us into 
believing, for example,  the phantom energy of the equation of state, $p=w\rho$ with $w<-1$, 
even in case that $w=-1$ is the true value.

\end{abstract}

\maketitle


\vskip1cm

\section{Introduction}\label{Sec1}

Usually, we believe that large scale isotropic inhomogeneities whose symmetry center coincide with our location   
are so small that we do not need to take into account them in the determination of cosmological parameters. 
This belief is based on the so called Copernican principle, which is one of the basic 
working hypotheses in the physical cosmology. 
Copernican principle states that we do not live in the privileged domain in the universe. 
If this is true, the isotropy around our location means that our universe is isotropic 
at every point, or equivalently, homogeneous and isotropic, and hence 
the observed high isotropy of the cosmic microwave background radiation 
strongly suggests the high homogeneity and isotropy of our universe in global sense 
(see, e.g. Ref.~\cite{wald}) which, at the same time, imply the smallness of the 
large scale inhomogeneities isotropic in terms of our location.

The universe model with large isotropic inhomogeneities have been studied in the context 
of the alternative scenario to explain the observed distance-redshift relation without 
introducing the dark energy components within the framework of general relativity. 
Although there are several severe observational constraints for the alternative 
scenario to the dark energy\cite{Tomita:2000jj,Tomita:2001gh,Celerier:1999hp,Iguchi:2001sq,Yoo:2008su,Bull:2012zx,Clifton:2008hv,Vanderveld:2006rb,Yoo:2010qn},   
we should note that this universe model has not been completely excluded yet. 
Furthermore, it should be noted that even if there are dark energy components, 
the existence of isotropic inhomogeneities may significantly affects observational results\cite{Romano:2010nc,Romano:2011mx,Marra:2010pg,Sinclair:2010sb,Valkenburg:2013qwa,Valkenburg:2012td,deLavallaz:2011tj,Valkenburg:2011ty,Marra:2012pj}.
Denoting the energy density and the pressure of the dark energy by $\rho_{\rm d}$ and $p_{\rm d}$, 
its equation of state is given by
\begin{equation}
p_{\rm d}=w\rho_{\rm d} \label{EOS}
\end{equation}
with $\rho_{\rm d}>0$ and $w<-1/3$. The special case of $w=-1$ corresponds to the cosmological constant. 
We would like to stress that if the observational data really implies that $w$ is less than $-1$, 
it is worthwhile to study the effects of  isotropic inhomogeneities
to the equation of state\cite{Valkenburg:2013qwa,Romano:2010nc,deLavallaz:2011tj,Valkenburg:2011ty,Marra:2012pj}; 
$w<-1$ may cause the causality violation. This is the main purpose of the present paper. 

In this paper, we focus on the effects of isotropic inhomogeneities on the estimate of how large amount of the 
dark energy in the universe and its equation of state. 
We consider the universe model filled with dust and a positive cosmological constant and 
assume that there are large-scale isotropic inhomogeneities which can be treated by the 
linear perturbation in the  Friedmann-Lema\^{\i}tre-Robertson-Walker (FLRW) universe model. 
This universe model has two arbitrary functions.  
In order to specify the two arbitrary functions, we need two conditions. 
One of these conditions is that we have considered only the growing mode of linear perturbation.
To specify one more function, we solve the inverse problem which is the reconstruction of the universe model from observational data.
We adopt distance-redshift relation as observational data. 
We solve the inverse problem with observational data which is the same as that of the universe filled with 
non-relativistic matter and dark energy with various $w$. 
To estimate the systematic error in determining the energy density and the equation of state of the 
dark energy due to isotropic inhomogeneities, we 
assume that inhomogeneities are described by the power spectrum of the random Gaussian statistics, which is 
consistent with the inflationary universe scenario. 
Then, we derive the probability that  
isotropic inhomogeneities obtained by solving the inverse problem are produced in the inflationary universe. 

The organization of this paper is as follows.
In Sec.~II, we derive the basic equations for inhomogeneous and isotropic universe model.
In Sec.~III, we derive the null geodesic equations in order to relate the distance-redshift relation 
with the universe model. In Sec.~IV, we derive the basic equations to determine the inhomogeneous and 
isotropic universe model from a given distance-redshift relation. 
We explain the numerical procedure in Sec.~V, and then we show the numerical results in Sec.~VI. 
We derive the probability of occurrence of systematic errors, in Sec.~VII.  
Finally, Sec.~VIII is devoted to the summary and discussion.

In this paper, we adopt the sign conventions of the metric and Riemann tensor of Ref.\cite{wald} and 
the geometrized unit in which the speed of light and Newton's gravitational constant are one. 

\section{Inhomogeneous isotropic universe model}\label{Sec2}

As mentioned in Sec.\ref{Sec1}, 
we consider the FLRW universe model with isotropic perturbations up to the linear order. 
Hereafter, we call this model the inhomogeneous and isotropic universe model. 
By adopting the Newtonian gauge (see e.g., \cite{Kodama-Sasaki}), 
the infinitesimal world interval is written in the form, 
\begin{equation}
ds^2=-\left[1+2\psi (t,\chi)\right]dt^2+a^2(t)\left[1-2\psi (t,\chi)\right]
\left[ d\chi^2+S^{2}(\chi;\bk)d\varOmega^2\right] ,   
\label{metric} 
\end{equation}
where $a(t)$ is the scale factor scaled so as to be unity at present time $t=t_0$, 
$d\varOmega^2$ is the line element of the unit 2-sphere, and, 
denoting the spatial Ricci curvature scalar of the background at present time by $6\bk$, the function 
$S(\chi;\bk)$ is defined as
\begin{equation}
S(\chi;\bk)=\frac{1}{\sqrt{\bk}}\sin \left(\sqrt{\bk}\chi\right).
\end{equation}
We assume that this universe model is filled with non-relativistic matter and the cosmological constant $\Lambda$, 
i.e., the so-called $\Lambda$CDM model. 
The stress-energy tensor of the non-relativistic matter, i.e., dust is given by
\begin{eqnarray}
T_{\mu\nu}=\bar{\rho}(1+\delta)\bar{u}_{\mu}\bar{u}_{\nu}
+\bar{\rho}\left(\bar{u}_{\mu}\delta u_{\nu}+\bar{u}_{\nu}\delta u_{\mu}\right), 
\label{T_ab} 
\end{eqnarray}
where $\bar{\rho}$ and $\bar{u}_\mu$ are the energy density and the 4-velocity of the background, 
respectively, whereas $\delta$ and $\delta u_\mu$ are the density contrast and the perturbation 
of the 4-velocity, respectively.
The coordinate system is chosen so that the components of the background velocity 4-vector 
and its perturbation are given by $\bar{u}_\mu=(-1,0,0,0)$ and $\delta u_\mu=(-\psi,av_{\chi},0,0)$.  

The Einstein equations lead to the Friedmann equation for the background;  
\begin{eqnarray}
\bH^{2}(a):=\left(\frac{1}{a}\frac{da}{dt}\right)^2=\frac{8\pi \brho _0}{3a^{3}} -\frac{\bk}{a^{2}}+\frac{\Lambda}{3} , 
\label{Eeq_b} 
\end{eqnarray}
where $\bar{\rho}_{0}$ is the background energy density at present time $t=t_0$. 
Denoting the present value of $\bH$ by $\bH_0$, the above equation is rewritten in the form
\begin{equation}
\bH^2=\bH_0^2\left(\frac{\Omega_{\rm m}}{a^3}+\frac{\Omega_{\rm k}}{a^2}+\Omega_\Lambda\right),
\label{H-eq}
\end{equation}
where
\begin{equation}
\Omega_{\rm m}=\frac{8\pi\bar{\rho}_{0}}{3\bH_0^2},~~~~\Omega_{\rm k}=-\frac{\bk}{\bH_0^2}~~~~{\rm and}~~~~
\Omega_\Lambda=\frac{\Lambda}{3\bH_0^2}.
\end{equation}

The Einstein equations lead to the equations for the linear perturbations;   
\begin{eqnarray}
\ddot{\psi} +4\bH\dot{\psi} +\left( \bH^{2}+2\frac{\ddot{a}}{a}-\frac{\bk}{a^{2}}\right) \psi=0,  
\label{Eeq_p1}
\end{eqnarray}
\begin{eqnarray}
\delta =\frac{a}{4\pi \brho_0}
\left[
\frac{1}{S^2}\partial_\chi
\left(S^2\partial_\chi \psi \right)
 +3\bk\psi 
 \right] -3f , 
\label{Eeq_p2}
\end{eqnarray}
\begin{eqnarray}
v_{\chi}=-\frac{a^{2}}{4\pi \bar{\rho}_{0}}\left(\partial _{\chi}\dot{\psi}+\bH\partial _{\chi}\psi \right) , 
\label{Eeq_p3}
\end{eqnarray}
where a dot denotes a partial differentiation with respect to $t$, 
and $f$ is the velocity potential which is related to the 
perturbation of 4-velocity through   
\begin{eqnarray}
v_{\chi}=-\frac{\partial _{\chi}f}{a\bH}. 
\label{v-f}
\end{eqnarray}

The general solution of Eq.~(\ref{Eeq_p1}) is represented 
by the linear superposition of the growing factor $D_{+}(t)$ and the decaying factor $D_{-}(t)$, which are defined as 
\begin{equation}
D_{+}(t):=\frac{\bH(a)}{a}\int ^{a}\frac{1}{b^{3}\bH^{3}(b)} db~~~~~{\rm and}~~~~~D_{-}(t):=\frac{\bH(a)}{a}.
\label{Dpm-sol}
\end{equation}
Hereafter, we assume that the decaying mode vanishes, since this assumption is consistent with 
the inflationary universe scenario.  
Thus, we have 
\begin{eqnarray}
\psi (t, \chi) =h(\chi) D_{+}(t) ,
\label{psi=hD}
\end{eqnarray}
where $h(\chi)$ is an arbitrary function of the radial coordinate $\chi$. 
Through Eqs.~(\ref{Eeq_p2}) and (\ref{Eeq_p3}), $\delta$ and $v_{\chi}$ are expressed 
by using $\psi$. Note that there are three parameters 
$\brho_0$, $\bk$, $\Lambda$ and one arbitrary function $h(\chi)$ 
in the inhomogeneous and isotropic universe model.

\section{Null geodesics in inhomogeneous and isotropic universe model}

Hereafter we drop higher-order terms in calculations without notifying that. 
All equations below are valid in the linear order of the perturbations.

We assume that the observer in the inhomogeneous and isotropic universe model 
stays at the symmetry center $\chi=0$ so that the observer recognizes 
the universe to be isotropic. In order to get the distance-redshift relation for the observer, 
we consider a past-directed 
radial null geodesic which emanates from the observer.  
By virtue of the isotropy in terms of the observer, both $k^{\theta}$ and $k^{\phi}$ should vanish.
One of the non-trivial components of the geodesic equations is given by
\begin{equation}
\frac{d}{d\lambda}\left[a^2(1-2hD_{+})k^\chi \right]+2a^2(D_{+}\partial _{\chi}h )\left(k^\chi\right)^2=0,
\label{g-eq-1}
\end{equation}
where $\lambda$ is the affine parameter. Equation (\ref{g-eq-1}) determines $k^\chi$, whereas 
the null condition determines $k^t$ in the manner
\begin{equation}
k^{t}=-a(1-2hD_{+})k^{\chi}.
\label{null-cond-1} 
\end{equation}
Then we have equations for $t$ and $\chi$ as 
\begin{eqnarray}
\frac{dt}{d\lambda}&=&k^t,  \label{g-eq-for-t} \\
\frac{d\chi}{d\lambda}&=&k^\chi. \label{g-eq-for-chi}
\end{eqnarray}

The redshift $z$ for the observer is given by  
\begin{equation}
1+z=\frac{(k^{\mu}u_{\mu})|_{{\rm s}}}{(k^{\mu}u_{\mu})|_{{\rm o}}} =-a\left(1- hD_+ +v_\chi\right)k^\chi,
\label{redshift-1}
\end{equation}
where subscripts s and o mean the quantities evaluated at the source and the observer, respectively, 
and we have chosen the affine parameter so that $-(k^\mu u_\mu)_{\rm o}$ is unity. 

We rewrite the equations for the radial null geodesic in the forms appropriate for later analyses.  
Equation~(\ref{g-eq-1}) is rewritten in the form, 
\begin{equation}
\frac{1}{k^\chi}\frac{d k^\chi}{dz}=-2\left(\bH-h\frac{dD_+}{dt}\right)\frac{dt}{dz}.
\label{g-eq-2}
\end{equation}
By differentiating the logarithm of each side of Eq.~(\ref{redshift-1}) with respect to $z$, we obtain
\begin{equation}
 \frac{1}{k^\chi}\frac{d k^\chi}{dz}=\frac{1}{1+z}-\left(\bH-h\frac{dD_+}{dt}\right)\frac{dt}{dz}+D_+\frac{dh}{dz}
 -\frac{dv_\chi}{dz}.
 \label{redshift-2}
 \end{equation}
From Eqs.~(\ref{g-eq-2}) and (\ref{redshift-2}), we have
\begin{equation}
\frac{1}{1+z}+\left(\bH-h\frac{dD_+}{dt}\right)\frac{dt}{dz}+D_+\frac{dh}{dz}-\frac{dv_\chi}{dz}=0.
\label{basic-eq-1}
\end{equation}
From the null condition (\ref{null-cond-1}), we have
\begin{equation}
\frac{dt}{dz}+a\left(1-2hD_+\right)\frac{d\chi}{dz}=0.
\label{basic-eq-2} 
\end{equation} 

We express the radial null geodesic as a function of $z$; 
\begin{eqnarray}
t&=&\bar{t}(z)+\delta t(z),  \label{t(z)} \\
\chi&=&\bar{\chi}(z)+\delta\chi(z), \label{chi(z)}
\end{eqnarray}
where the quantities with a horizontal bar represent the background solution. 
From Eqs.~(\ref{basic-eq-1}) and (\ref{basic-eq-2}), we see that 
the background solutions $\bar{t}(z)$ and $\bar{\chi}(z)$ satisfy  
\begin{eqnarray}
\frac{d\bar{t}}{dz}=-\frac{1}{(1+z)\bar{H}}, 
\label{g-eq_bt}
\end{eqnarray}
\begin{equation}
\frac{d\bar{\chi}}{dz}=\frac{1}{\bar{H}}.
\label{g-eq_bchi}
\end{equation}
Here note that 
\begin{eqnarray}
\bar{H}=\bH_{0}\sqrt{\Omega _{\rm m}(1+z)^{3}+\Omega _{\rm k}(1+z)^{2}+\Omega _{\Lambda}}~. 
\label{Eeq_bz} 
\end{eqnarray}
Equation~(\ref{basic-eq-1}) leads to  the equation for the linear perturbations as
\begin{equation}
\bar{H}\frac{d\delta t}{dz}+\frac{d\bar{H}}{dz}\delta t-\frac{dD_{+}}{dz}h+D_{+}\frac{dh}{dz}-\frac{dv_\chi}{dz}=0,
\label{basic-1}
\end{equation}
whereas Eq.~(\ref{basic-eq-2}) leads to 
\begin{equation}
\frac{d\delta t}{dz}+\frac{1}{1+z}\left(\frac{d\delta\chi}{dz}+\delta t-\frac{2D_{+}h}{\bar{H}}\right)=0,
\label{basic-2}
\end{equation}
where we have used Eqs.~(\ref{g-eq_bt}) and (\ref{g-eq_bchi}). 
If the inhomogeneous and isotropic universe model is completely fixed, 
Eqs.~(\ref{basic-1}) and (\ref{basic-2}) determine $\delta t$ and $\delta\chi$. 

The boundary conditions for Eqs.~(\ref{g-eq_bt}), (\ref{g-eq_bchi}), (\ref{basic-1}) and (\ref{basic-2}) 
are given by 
\begin{equation}
\bar{t}(0)=t_0~~~~{\rm and}~~~~\bar{\chi}(0)=\delta t (0)=\delta \chi (0)=0. \label{b-condi} 
\end{equation}

Note that $h$, $D_+$ and $v_\chi$ are related to each other through Eq.~(\ref{Eeq_p3}) in the manner
\begin{eqnarray}
v_\chi-\frac{2\bar{H}^{2}}{3\bH_{0}^{2}\Omega _{\rm m}(1+z)^{2}}
\left[ (1+z)\frac{dD_{+}}{dz} -D_{+} \right]\frac{dh}{dz}=0. 
\label{basic-3}
\end{eqnarray}

\section{The effects of inhomogeneities on the estimate of dark energy}

In cosmology, the observable domain is restricted mainly to the null hypersurface 
generated by past directed null geodesics emanated from the observer. 
This restriction causes the difficulty in recognizing isotropic  
inhomogeneities, since it is extremely difficult without changing the observation site   
over the cosmological scale to check whether such inhomogeneities exist.  
As mentioned, we study the possibility that the effects of these formidable isotropic inhomogeneities  
and the functional degree of freedom of $w$ in the equation of state of the 
dark energy (\ref{EOS}) degenerate. 

We assume that the hypothetical observational data determines the dependence of   
the angular diameter distance $d_{\rm A}$ on the redshift $z$ in the form 
\begin{equation}
d_{\rm A}=D_{\rm A}(z). \label{data}
\end{equation}
Hereafter, we focus on the situation in which if we assume general relativity and the 
FLRW universe model, the observational data (\ref{data}) 
indicates that the universe is dominated  
by the non-relativistic matter and the dark energy 
whose $w$ in the equation of state (\ref{EOS}) is  written in the form
\begin{eqnarray}
w=\sum_{n=0}^\infty w_n(1-a)^n. \label{w-def} 
\end{eqnarray}
where $a$ is the scale factor, and $w_n$ is constant and $w_0 < -1/3$.  
We assume that $w_n=0$ for $n\geq2$. 
Furthermore, for simplicity, we assume that 
the observationally indicated FLRW universe model 
has flat space, $\bk=0$. 
This FLRW model is characterized by four parameters, Hubble constant $H_0$,  
$w_{0}$, $w_1$ and the present value of the energy density $\rho_{\rm d0}$ of the dark energy, 
or equivalently, 
\begin{equation}
\Omega _{\rm d}:=\frac{8\pi \rho_{\rm d0}}{3H_0^2}.
\end{equation}

In the case of the real observation, 
the parameters $H_0$, $w_0$, $w_1$ and $\Omega_{\rm d}$ 
are determined so that the observed distance-redshift relation (\ref{data}) is well fitted.  
However, the present analysis is theoretical, and hence 
instead of fixing the functional form of hypothetical observational data $D_{\rm A}(z)$, 
we determine the cosmological parameters $H_0$, $w_0$, $w_1$ and $\Omega_{\rm d}$ first. 
Then we obtain the angular diameter distance of this FLRW universe model  
and identify it with $D_{\rm A}(z)$.

As mentioned in Sec.~\ref{Sec1}, 
our purpose is to study whether the hypothetical observational data (\ref{data}) can also be 
explained by the inhomogeneous and isotropic universe model given in Sec.~\ref{Sec2} 
within the framework of the linear perturbation theory. 
In this case, the basic equations for the radial null geodesic (\ref{basic-1}) and (\ref{basic-2}) and 
(\ref{basic-3}) should be regarded as the system of differential equations to determine the 
inhomogeneous and isotropic universe model, i.e., the parameters 
$\bH_0$, $\Omega_{\rm m}$, $\Omega_{\rm k}$, $\Omega_\Lambda$ and 
the functional degree of freedom $h(\chi)$; We will rewrite them into the forms appropriate 
for this purpose. 

In the inhomogeneous and isotropic universe model, 
the angular diameter distance from some light source to the central observer 
is equal to the areal radius at which the light is emitted; 
\begin{eqnarray}
d_{\rm A}(z)&=&a(t)\Bigl[1-D_{+}(t)h(\chi)\Bigr]S(\chi;\bk)\Bigl|_{t=t(z),~\chi=\chi(z)} \cr
&&\cr
&=&a(\bar{t})S(\bar{\chi};\bk)+a(\bar{t})\left[(\bar{H}\delta t-D_{+}h)S(\bar{\chi};\bk)
+S'(\bar{\chi};\bk) \delta\chi\right] ,
\label{condi-1}
\end{eqnarray}
where a prime means a derivative with respect to its argument. 
Then, the angular diameter distance (\ref{condi-1}) is assumed to satisfy Eq.~(\ref{data}). 
The angular diameter distance in the background FLRW 
universe model $\bar{d}_{\rm A}(z)$ is given by
\begin{equation}
\bar{d}_{\rm A}(z)=\frac{S(\bar{\chi};\bk)}{1+z}.
\end{equation}
We define
\begin{equation}
\delta d_{\rm A}(z) =D_{\rm A}(z)-\bar{d}_{\rm A}(z).
\end{equation}
Then, Eq.~(\ref{condi-1}) leads to  
\begin{eqnarray}
\delta \chi(z)=\frac{(1+z)\delta d_{\rm A}(z)
-\left[\bar{H}(z)\delta t (z)-D_{+}(\bar{t})h(\bar{\chi})\right]S(\bar{\chi};\bk)}{S'(\bar{\chi};\bk)}~. 
\label{condi-2}
\end{eqnarray}
Equation (\ref{condi-2}) is equivalent to the hypothetical observed distance-redshift relation (\ref{data}). 

We regard Eq.~(\ref{basic-3}) as the differential equation for $h$. 
By substituting Eq.~(\ref{condi-2}) into Eq.~(\ref{basic-2}), we eliminate $\delta\chi$ from Eq.~(\ref{basic-2}) 
and obtain the differential equation for $\delta t$.  
In order to eliminate $d\delta t/dz$ and $dh/dz$ from Eq.~(\ref{basic-1}), we use Eq.~(\ref{basic-2}) 
with $\delta\chi$ eliminated and Eq.~(\ref{basic-3}), and then, we obtain the differential equation for $v_\chi$. 
As a result, we obtain the following system of differential equations to determine the 
inhomogeneous and isotropic universe model from the hypothetical observed distance-redshift relation 
(\ref{data});  
\begin{eqnarray}
\frac{d h}{dz}&=& {\cal H}\left(v_\chi, z\right), 
\label{eq-1} \\
&& \cr
\frac{d\delta t}{dz}&=&
\frac{{\cal N}\left(h,\delta t,v_{\chi},z\right)}{{\cal D}(z)},
\label{eq-2} \\
&&\cr
\frac{d v_\chi}{dz}&=& {\cal V}\left(
h,\delta t,v_\chi, z\right),
\label{eq-3} 
\end{eqnarray}
where
\begin{eqnarray}
{\cal H}&=&-\frac{3\bH_{0}^{2}\Omega _{\rm m}(1+z)^{2}v_\chi}{2\bar{H}^{2}}\left[D_{+}- (1+z)\frac{dD_{+}}{dz} \right]^{-1},\\
&&\cr
{\cal D}(z)
&=&
1-\frac{\bar{H}}{(1+z)}\frac{S(\bar{\chi};\bk)}{S'(\bar{\chi};\bk)}, \\
&&\cr
{\cal N}\left(h,\delta t,v_{\chi},z\right)
&=&
\frac{1}{1+z}\left[
-\frac{d}{dz}\left(\frac{(1+z)\delta d_A}{S'}\right)
+\left\{-1+\frac{d}{dz}\left(\frac{S\bH}{S'}\right)\right\}\delta t \right. \cr
&& \cr
& &{}  
\left.
+\left\{\frac{2D_+}{\bH}-\frac{d}{dz}\left(\frac{D_+S}{S'}\right)\right\}h
-\frac{D_+S}{S'}{\cal H}\right]
 \\
&& \cr 
{\cal V}&=&
-\frac{dD_{+}}{dz}h+D_{+}{\cal H}
+\frac{d\bar{H}}{dz}\delta t
+\bar{H}\frac{{\cal N}}{{\cal D}}.
\end{eqnarray}

The boundary conditions for Eqs.~(\ref{eq-1})--(\ref{eq-3}) are given as follows. 
The value of $h|_{z=0}$ can be made zero by the rescaling of the coordinate and introduce a new scale factor. 
(see Appendix A). 
Hence we impose  \begin{eqnarray}
h|_{z=0}=0. 
\label{bound_h} 
\end{eqnarray} 
The boundary condition on $\delta t(z)$ is given by Eq.~(\ref{b-condi}). 
Imposing $C^1$ regularity for $h$ at the origin, that is, $\partial_\chi h |_{z=0}=0$, 
from Eq.~(\ref{Eeq_p3}), we obtain
\begin{eqnarray}
v_\chi|_{z=0}=0.  
\label{bound_v} 
\end{eqnarray} 
We numerically solve Eqs.~(\ref{eq-1})--(\ref{eq-3}) under the boundary conditions 
(\ref{b-condi}), (\ref{bound_h}) and (\ref{bound_v}). 
If we find numerical solutions are so small that the linear approximation is valid, 
we may say that the FLRW universe model filled with 
the dark energy whose equation of state is given by Eq.~(\ref{w-def}) with a set of $w_n$ 
can also be explained by the inhomogeneities and isotropic universe model with the dark energy of 
$w_0=-1$ and $w_n=0$ of $n\geq1$, i.e., the cosmological constant. 

\section{numerical procedure}

Before numerically integrating Eqs.~(\ref{eq-1})--(\ref{eq-3}),  
we need to fix three cosmological 
parameters $\bH_{0}$, $\Omega _{\rm k}$ and $\Omega _{\Lambda}$ of the 
inhomogeneous and isotropic universe model. 
The Hubble parameter $\bH_{0}$ should be the 
same as that of the FLRW universe model giving $D_{\rm A}(z)$, 
since it is an observable quantity. 
By contrast, as explained below, one of $\Omega_{\rm k}$ and $\Omega_\Lambda$ cannot be 
freely determined as long as we demand that the solutions are at least everywhere $C^1$.  

Equation~(\ref{eq-2}) shows that this equation has a regular singular point at 
$z=z_{\rm cr}$ which is a root of ${\cal D}(z)=0$.  
The function ${\cal N}(z):={\cal N}(h(z),\delta t(z), v_\chi(z),z)$ should satisfy
\begin{equation}
{\cal N}(z_{\rm cr})=0 \label{regularity}
\end{equation}
so that the solutions of Eqs.~(\ref{eq-1})--(\ref{eq-3}) have finite and continuous 
derivatives with respect to $z$ at $z=z_{\rm cr}$. The condition (\ref{regularity}) leads to 
a relation between $\Omega_{\rm k}$ and $\Omega_\Lambda$. 
In this paper, we fix $\Omega_\Lambda=0.7$, and then we search for 
$\Omega_{\rm k}$ which guarantees the smoothness of the solutions at $z=z_{\rm cr}$ by 
a kind of the shooting method as shown below. 

Since we can find $z_{\rm cr}>1$ for the cases of our interest, we 
solve Eqs.~(\ref{eq-1})--(\ref{eq-3}) from $z=0$ to $z=1$ by imposing 
the boundary conditions (\ref{b-condi}), (\ref{bound_h}) and (\ref{bound_v}) and, 
at the same time, from $z=z_{\rm cr}$ to $z=1$, by making a guess at    
$\Omega_{\rm k}$, $h|_{z=z_{\rm cr}}$ and $v_\chi|_{z=z_{\rm cr}}$ and then fixing   
$\delta t(z_{\rm cr})$ so that Eq.~(\ref{regularity}) is satisfied. 
If we fail to get smooth solutions of Eqs.~(\ref{eq-1})--(\ref{eq-3}), we select different 
values of $\Omega_{\rm k}$, $h|_{z=z_{\rm cr}}$ and $v_\chi|_{z=z_{\rm cr}}$ in accordance 
with the Newton method in the three-dimensional parameter space 
and then again integrate Eqs.~(\ref{eq-1})--(\ref{eq-3}) from 
$z=0$ to $z=1$ and, at the same time, from  $z=z_{\rm cr}$ to $z=1$. 
We iterate this procedure until we obtain numerical solutions sufficiently smooth at $z=1$. 
Next, we integrate Eqs.~(\ref{eq-1})--(\ref{eq-3}) outward from $z=z_{\rm cr}$ 
with the values of $\Omega_{\rm k}$,  $h|_{z=z_{\rm cr}}$ and $v_\chi|_{z=z_{\rm cr}}$ 
which guarantee the smoothness of the solutions.

\section{numerical results}

Here, we introduce the density contrast in the synchronous comoving (SC)  
gauge $\Delta$ defined as
\begin{eqnarray}
\Delta := \delta +3f.
 \label{Delta-def}
\end{eqnarray} 
where $f$ is the velocity potential introduced in Eqs~(\ref{Eeq_p2}) and (\ref{v-f}). 
The gauge transformation from the Newtonian gauge to the SC gauge is given in Appendix B. 

At present, we do not have any observational data of the distance-redshift 
relation in the domain of $z\geq 2$, except for positions of the acoustic peaks in the spectrum of 
Cosmic Microwave Background Radiation (CMBR)\cite{YNS2010}. 
Hence we assume that the universe model obtained by solving the inverse problem is 
available in the domain of $z<2$ only, and the domain of $z > z_{\rm b}$ ($z_{\rm b}>2$)  
agrees with the FLRW universe model with $\bk=0$, i.e., the concordance $\Lambda$CDM model, 
since the concordance $\Lambda$CDM model is consistent with many of inflationary universe models. 
We call this concordance $\Lambda$CDM universe model the faraway background. 
It should be noted that the faraway background 
does not necessarily agree with the background universe model introduced in the previous section, 
and hence we have to introduce the density contrast relative to the faraway background.

We assume that the present energy density satisfies 
\begin{equation}
 \int_0^{r_{\rm b}} \fDelta(\tau_0,r)r^2dr=0,
 \label{conp}
\end{equation}
where $\fDelta(\tau_0,r)$ is the density contrast relative to the faraway background 
at the present SC time $\tau=\tau_0$, and 
$r_{\rm b}$ is the comoving radial coordinate at which $z=z_{\rm b}$ (see Appendix C on 
the SC coordinates $\tau$ and $r$).  
The condition (\ref{conp}) is necessary so that the domain of $r\geq r_{\rm b}$ is exactly 
the concordance $\Lambda$CDM model. But we do not impose any additional conditions 
on $\fDelta$ of $2\leq z <z_{\rm b}$ besides Eq.~(\ref{conp}).  

\begin{figure}[htbp]
 \begin{center}
 \includegraphics[width=100mm]{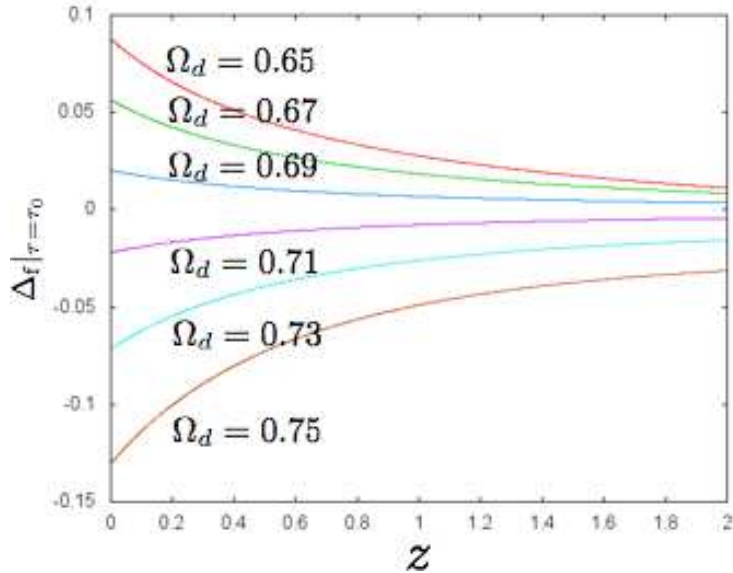}
 \end{center}
 \caption{
 We depict the present gauge invariant density contrast $\fDelta|_{\tau=\tau_0}$ 
 in the homogeneous and isotropic background with $\Omega_\Lambda=0.7$ 
 as a function of the redshift $z$, which can explain the distance-redshift relation 
 of the FLRW universe model with 
 various $\Omega_{\rm d}$ of the dark energy with $w_0=-1$ and $w_1=0$. 
  }
 \label{fig:one}
\end{figure}

\begin{figure}[htbp]
 \begin{center}
\includegraphics[width=100mm]{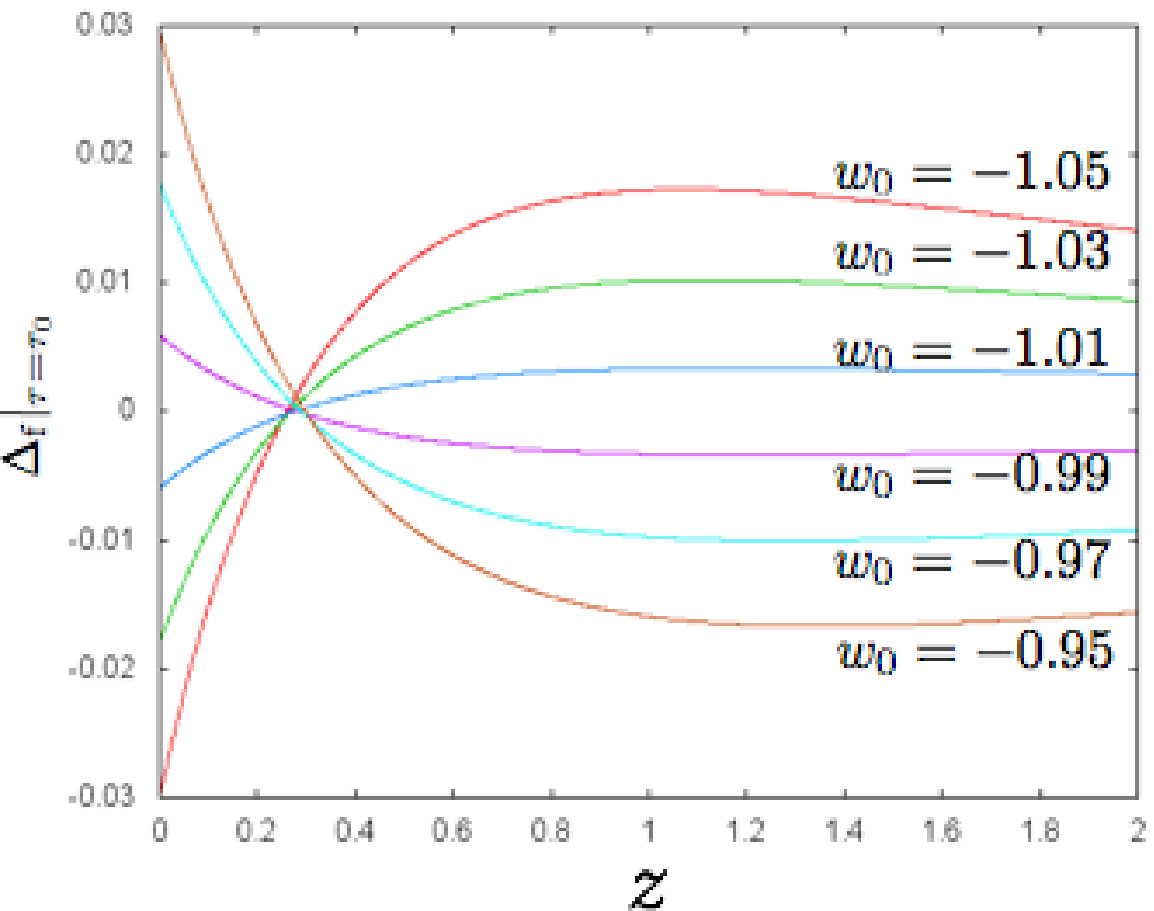}
 \end{center}
 \caption{The same as Fig. \ref{fig:one}, but $\Omega _{\rm d}=0.7$ and $w_1=0$ for various $w_0$. }
 \label{fig:two}
\end{figure}

\begin{figure}[htbp]
 \begin{center}
\includegraphics[width=100mm]{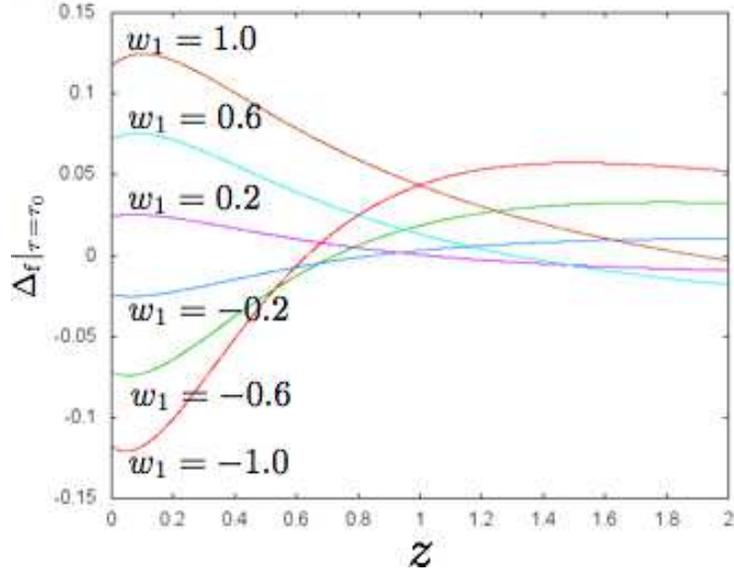}
 \end{center}
 \caption{The same as Fig. \ref{fig:one}, but $\Omega _{\rm d}=0.7$ and $w_0=-1$ for various $w_1$. }
 \label{fig:three}
\end{figure}

In Figs.~\ref{fig:one}--\ref{fig:three}, 
we depict the present density contrast $\fDelta|_{\tau=\tau_0}$ in the 
homogeneous and isotropic background with $\Omega_\Lambda=0.7$ as a function of the redshifht $z$,  
which can explain the angular diameter distance agrees with $D_{\rm A}(z)$ 
of the FLRW universe model with the dark energy of various $\Omega_{\rm d}$,  
$w_0$ and $w_1$.  The outermost data points correspond to that of $z=2$.

In Fig.~\ref{fig:one}, we depict $\fDelta|_{\tau=\tau_0}$ 
in the case of $w_0=-1$, $w_1=0$ and various $\Omega_{\rm d}$.  
In Fig.~\ref{fig:two}, we depict the same as Fig.~\ref{fig:one} but in the case of 
$\Omega _{\rm d}=0.7$ and $w_1=0$ with various $w_0$.  
In Fig.~\ref{fig:three}, we depict the same as Fig.~\ref{fig:one} but in the case of 
$\Omega _{\rm d}=0.7$ and $w_0=-1$ with various $w_1$. 
We can see form these figures 
that the isotropic inhomogeneities may mislead us about the estimate of 
the values of $\Omega_{\rm k}$, $w_0$ and $w_1$.

\section{Probability of occurrence}

We have seen that the isotropic perturbations can cause systematic errors in the estimates of the 
cosmological parameters and the equation of state of the dark energy. 

In this section, we show the probability of occurrence of these isotropic perturbations. 
We assume that the initial density perturbations obey random Gaussian statistics which is consistent to 
the inflationary universe scenario:  
the Fourier transform of the density contrast is given by
\begin{equation}
\tilde{\Delta}(\tau,\bm{k})=\int d^3x~\Delta(\tau,\bm{x})e^{-i\bm{k}\cdot\bm{x}}, 
\end{equation}
and then, we have
\begin{equation} 
\langle\tilde{\Delta}(t,\bm{k})\tilde{\Delta}^*(t,\bm{k}')\rangle
=(2\pi)^3\delta^3_{\rm D}(\bm{k}-\bm{k}')P(t,k),
\end{equation}
where $\langle \cdot\cdot\cdot\rangle$ represents the ensemble mean, 
$\delta_{\rm D}(x)$ is Dirac's delta function, and $P(t,k)$ is the power spectrum at time $t$. 
We assume 
\begin{equation}
P(t,k)=A_0 k^n T^2(k)D_+^2(t)
\end{equation}
where $n$ and $A_0$ are constants determined by observing the 
fluctuation of the CMBR, whereas $T(k)$ is the matter transfer function; We adopt 
$n$ and $A_0$ derived from the Planck data 
and the matter transfer function derived by Eisenstein and Hu\cite{Eisenstein:1997jh}. 
The matter transfer function is determined by background cosmological parameters.

The mass contrast within the radius $r$ is given by
\begin{equation}
\frac{\delta M}{M}(\tau,r)=\frac{1}{V}\int d^3x~ \vartheta\left(r-|\bm{x}|\right)
\Delta(t,\bm{x}),
\end{equation}
where $\vartheta(z)$ is Heaviside's step function, and
\begin{equation}
V=\int d^3x~ \vartheta\left(r-|\bm{x}|\right)=\frac{4}{3}\pi r^3.
\end{equation}
The mean square of the mass contrast is then given by
\begin{equation}
\sigma^2(\tau,r):=\left\langle \left(\frac{\delta M}{M}(\tau,r)\right)^2  \right\rangle
=\frac{1}{2\pi^2}\int_0^\infty P(t,k)W(kr)k^2dk,
\end{equation}
where
\begin{equation}
W(y)=\frac{9}{y^6}(\sin y-y\cos y)^2.
\end{equation}

\begin{figure}[htbp]
 \begin{center}
\includegraphics[width=100mm]{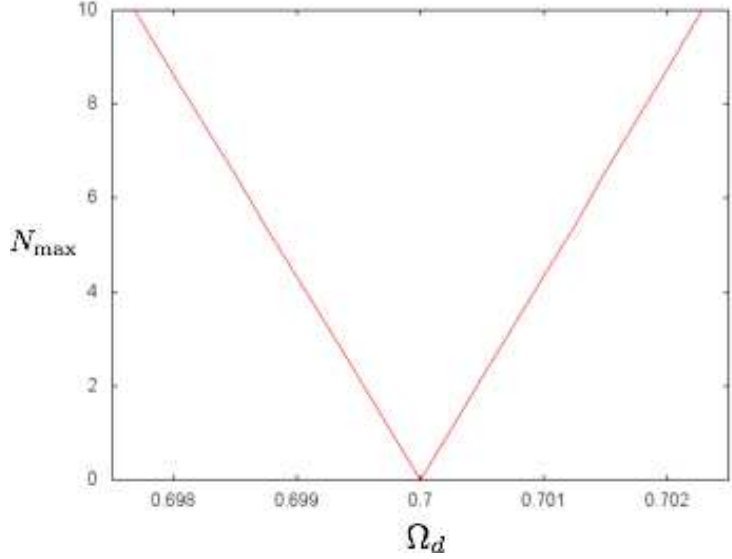}
 \end{center}
 \caption{$N_{\rm max}$ is depicted as a function of $\Omega_{\rm d}$ 
 We fix the parameters of the FLRW universe with dark energy as $\Omega _\Lambda=0.7$,  
 $w_{0}=-1.0$ and $w_1=0$. }
 \label{fig:four}
\end{figure}

\begin{figure}[htbp]
 \begin{center}
 \includegraphics[width=100mm]{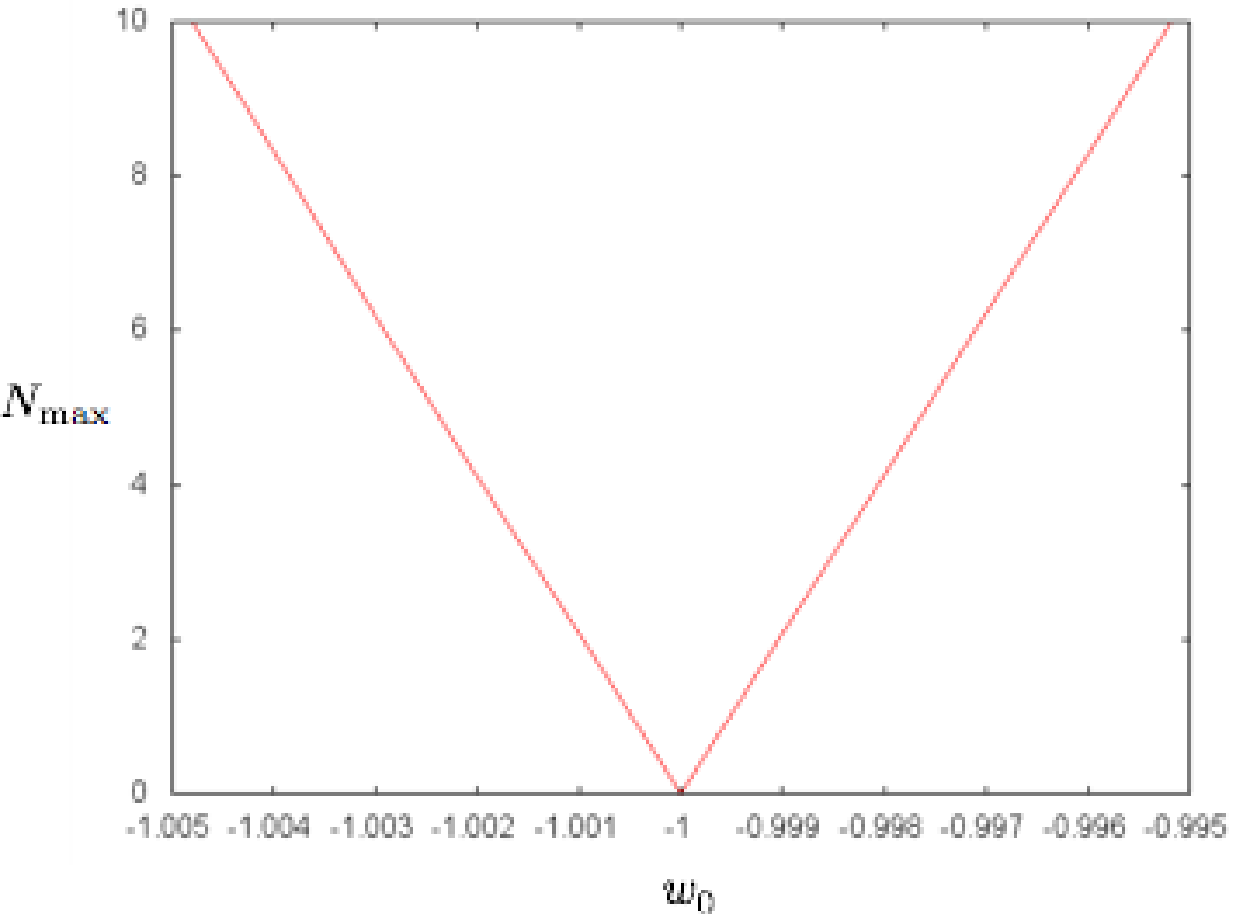}
 \end{center}
 \caption{The same as Fig. \ref{fig:four}, but $\Omega _{\rm d}=0.7$ and $w_1=0$ for various $\Delta w_0$. }
 \label{fig:five}
\end{figure}

\begin{figure}[htbp]
 \begin{center}
\includegraphics[width=100mm]{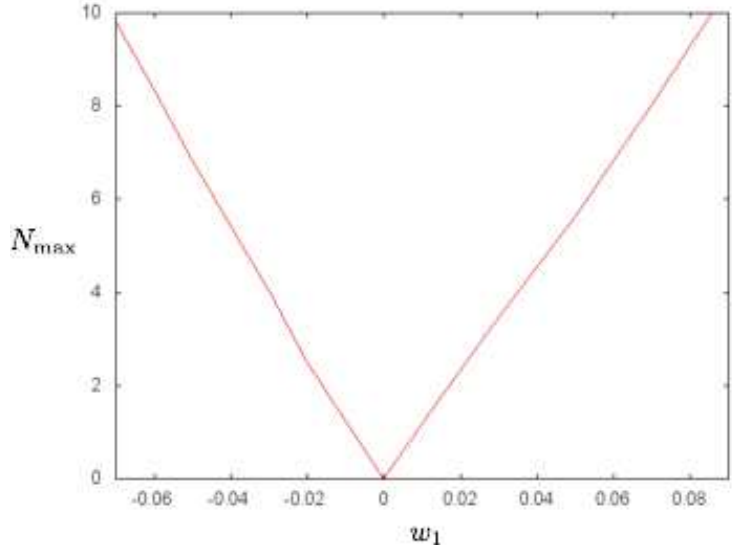}
 \end{center}
 \caption{The same as Fig. \ref{fig:four}, but $\Omega _{\rm d}=0.7$ and $w_0=-1$ for various $w_1$. }
 \label{fig:six}
\end{figure}

The present mass contrast of the inhomogeneous and isotropic universe model obtained by solving the 
inverse problem 
is defined as
\begin{eqnarray}
\Delta_{\rm M}(r):&=&\left| \frac{4\pi}{V} \int_0^r \fDelta(\tau_0,y) y^2 dy\right|. 
\end{eqnarray}
Then, we define 
\begin{equation}
N(r) :=\frac{\Delta_{\rm  M}(r)}{\sigma(\tau_0, r)}. 
\end{equation}
The larger $N(r)$ implies that such an inhomogeneity is rare. The maximal value of $N(r)$ in 
$0<r<r_2$ is denoted by $N_{\rm max}$, where $r=r_2$ is the comoving radial coordinate of $z=2$. 

In Figs.~\ref{fig:four}--\ref{fig:six}, we depict the $N_{\rm max}$ 
with the dark energy of various $\Omega_{\rm d}$, $w_0$ and $w_1$.
In Fig.~\ref{fig:four}, we depict $N_{\rm max}$ as a function of $\Omega_{\rm d}$ 
in the case of $w_0=-1$, $w_1=0$. 
$\Omega_{\rm d}$ close to $0.7$ means lower density contrast.
In Fig.~\ref{fig:five}, we depict the same as Fig.~\ref{fig:four} but in the case of 
$\Omega _{\rm d}=0.7$ and $w_1=0$ with various $w_0$.
 In Fig.\ref{fig:six}, we depict the same as Fig.~\ref{fig:four} but in the case of 
$\Omega _{\rm d}=0.7$ and $w_0=-1$ with various $w_1$.

\section{Summary and discussion}

We studied the systematic error caused by isotropic inhomogeneities, which 
appear in determining the amount of the dark energy and 
its equation of state.  
We have provided the hypothetical observational data which are equivalent to those 
obtained in the universe dominated by the dark energy of various $\Omega_{\rm d}$ and 
equation of state. Then we solved the inverse problem to evaluate the density contrast 
in the FLRW background filled with non-relativistic matter and the cosmological constant, 
so that the hypothetical observational data is explained. 
The probability of occurrence of these isotropic perturbations is also 
evaluated in accordance with the standard inflationary universe scenario together with the 
Planck data.  

Our result implies that the uncertainty in $w_0$ is $\pm4\times10^{-3}$ in 10$\sigma$ confidence level, and  
even if we get the estimate $w_0=-1.005$, it does not necessarily mean the existence of the phantom 
energy. Even in the case that observational data suggest $w_0=-1.01$,  it is impossible to 
deny the possibility that it comes from a large scale isotropic inhomogeneity;  
since it may be only one realization within our observable domain, we can not exclude,  
from the statistical argument, the possibility that such an inhomogeneity unexpectedly appears. 
Although it is very challenging to observationally determine how large the isotropic perturbations are,  
we need constraints on their existence in order to arrive at the 
conclusion with sufficiently confidence. This is the future problem.

\section*{Acknowledgments}
We are grateful to Hideki Ishihara and colleagues in the group of elementary 
particle physics and gravity at Osaka City University for useful discussions and helpful comments.  
C-MY thanks Masato Tokutake for his useful comments. 
KN was supported in part by JSPS Grant-in-Aid for Scientific Research (C) (No. 25400265).

\appendix
\section{Rescaling to eliminate the metric perturbation at the symmetry center}

In this paper, we consider the isotropically perturbed 
FLRW universe model filled with non-relativistic matter and the cosmological constant. As shown in 
Sec.~\ref{Sec2}, the metric is 
given by 
\begin{equation}
ds^2=-\left[1+2h(\chi)D_+(t)\right]dt^2
+a^2(t)\left[1-2h(\chi)D_+(t)\right]\left[d\chi^2+S^2(\chi;\bk)d\varOmega^2\right].
\label{metric-A1}
\end{equation}
We rescale the time and radial coordinates in the manner
\begin{eqnarray}
d\check{t}&=&\left[1+h(0)D_{+}(t)\right]dt, \\
\label{new-t}
\check{\chi}&=&\left[1-h(0)D_+(t_0)\right]\chi, \label{new-chi}
\end{eqnarray}
and introduce a new scale factor defined as
\begin{eqnarray}
\check{a}(\check{t})=\left[1-h(0)\left\{D_{+}(t)-D_{+}(t_0)\right\}\right]a(t).
\label{tilde-a}
\end{eqnarray}
Then, up to the first order perturbations, the line element is rewritten in the form 
\begin{equation}
ds^2=-\left[1+2\check{h}(\check{\chi})\check{D}_+(\check{t})\right]d\check{t}^2
+\check{a}^2(\check{t})\left[1-2\check{h}(\check{\chi})\check{D}_+(\check{t})\right]\left[d\check{\chi}^2
+S^2(\check{\chi};\bk)d\varOmega^2\right],
\label{metric-A2}
\end{equation}
where
\begin{eqnarray}
\check{k}&=&\left[1+2h(0)D_{+}(t_0)\right]\bk \label{tilde-k}, \\
\check{h}(\check{\chi})&=&h(\chi)-h(0), \label{tilde-h} \\
\check{D}_+(\check{t})&=&D_+(t). \label{tilde-D} 
\end{eqnarray}
Note that $\check{h}$ vanishes at the symmetry center $\check{\chi}=0=\chi$. 

The Hubble equation for the new scale factor $\check{a}$ is given by
\begin{eqnarray}
\left(\frac{1}{\check{a}(\check{t})}\frac{d\check{a}(\check{t})}{d\check{t}}\right)^2
&=&\left[1-2h(0)D_+(t)\right]\left(\frac{1}{a(t)}\frac{da(t)}{dt}\right)^2
-2h(0)\left(\frac{1}{a(t)}\frac{da(t)}{dt}\right)
\frac{dD_+(t)}{dt} \cr
&=&\bH_0^2\left[\frac{\Omega_{\rm m}}{a^3(t)}
\left(1+3h(0)D_+(t)-\frac{2h(0)}{\Omega_{\rm m}\bH_0^2}\right)
+\left[1+2h(0)D_+(t)\right]\frac{\Omega_{\rm k}}{a^2(t)}+\Omega_\Lambda\right] \cr
&=&\bH_0^2\left(\frac{\check{\Omega}_{\rm m}}{\check{a}^3(\check{t})}
+\frac{\check{\Omega}_{\rm k}}{\check{a}^2(\check{t})}
+\Omega_{\Lambda}\right), \label{H-eq-2}
\end{eqnarray}
where we used 
the original background Hubble equation (\ref{H-eq}) and Eq.~(\ref{Dpm-sol}) and furthermore 
introduced the following new cosmological parameters:
\begin{eqnarray}
\check{\Omega}_{\rm m}&=&\Omega_{\rm m}
\left[1+3h(0)D_+(t_0)-\frac{2h(0)}{\Omega_{\rm m}\bH_0^2}\right], \\
\check{\Omega}_{\rm k}&=&\Omega_{\rm k}\left[1+2h(0)D_+(t_0)\right]. 
\end{eqnarray}
Equation (\ref{H-eq-2}) implies that 
$\check{a}$ is regarded as the scale factor of the FLRW universe model of the cosmological 
parameters $\check{\Omega}_{\rm m}$, $\check{\Omega}_{\rm k}$ and $\Omega_\Lambda$. 
Hence, without loss of generality, we can get the isotropic metric perturbations to 
vanish at the symmetry center $\chi=0$ by choosing an appropriate background 
FLRW universe. 

\section{Gauge transformation}

The transformation between the Newton and synchronous comoving (SC) gauges is given by
\begin{eqnarray}
d\tau&=&(1+\psi)dt-av_\chi d\chi, \label{dtau}\\
dx&=&-\frac{v_\chi}{a}dt+\left(1+\frac{a}{4\pi\bar{\rho}_0}\partial_\chi\psi\right)d\chi. \label{dx}
\end{eqnarray}
By integrating the above equations, we have
\begin{eqnarray}
\tau&=&t+\frac{f}{H}, \\
x&=&\chi+\frac{a}{4\pi\bar{\rho}_0}\partial_\chi\psi.  
\end{eqnarray}
where we have used Eq.~(\ref{Eeq_p3}) in obtaining $r$, whereas Eq.~(\ref{v-f}) 
has been used in obtaining $\tau$. 
The infinitesimal world interval in the SC gauge is given by
\begin{eqnarray}
&&ds^2=-d\tau^2+a^2(\tau)\left[1-2(\psi+f)\right]\cr
&&\cr
&&\times\left[\left(1-\frac{a}{2\pi\brho_0}\partial_x^2\psi\right)dx^2
+S^2(x;\bk)\left(1-\frac{a}{2\pi\brho_0}
\frac{S'(x;\bk)}{S(r;\bk)}\partial_x\psi\right)d\varOmega^2\right]. 
\label{SC-gauge}
\end{eqnarray}
By the transformation (\ref{dtau}) and (\ref{dx}), we have, up to the first order,  
\begin{eqnarray}
T_{\tau\tau}&=&\left(\frac{\partial t}{\partial \tau}\right)_x^2T_{tt}
+2\left(\frac{\partial t}{\partial \tau}\right)_x\left(\frac{\partial \chi}{\partial \tau}\right)_xT_{t\chi}
+\left(\frac{\partial \chi}{\partial \tau}\right)_x^2T_{\chi\chi} 
=\bar{\rho}(\tau)\left(1+\delta+3f\right),
\end{eqnarray}
where $T_{tt}=\brho(t)(1+\delta+2\psi)$ has been used. Then, since 
$T_{\tau\tau}=\brho(\tau)(1+\Delta)$, we have Eq.~(\ref{Delta-def}). 

\section{Replacement of the background}

Here, we consider the replacement of the background from the FLRW 
universe model of $\bk\neq0$ 
to that of $\bk=0$. The FLRW universe model is assumed to 
be filled with the dust and the cosmological 
constant. 

We write the FLRW universe model with $\bk\neq0$ in the form 
\begin{equation}
ds^2=-d\tau^2+a^2(\tau)\left(1+\frac{\bk}{4}R^2\right)^{-2}
\left(dR^2+R^2d\Omega^2\right).
\label{LCDM}
\end{equation}
The relation between the radial coordinates $\chi$ in Eq.~(\ref{metric}) and $R$ is given by
\begin{equation}
R=\frac{2}{\bk S(x;\bk)}\left[1-\sqrt{1-\bk S^2(x;\bk)}\right].
\end{equation}

We assume $|\bk R^2|\lesssim |k/\brho_0|\ll1$. Then, Eq.~(\ref{LCDM}) is rewritten in the form of the 
infinitesimal world interval of the FLRW of $\bk=0$ with perturbations as
\begin{equation}
ds^2=-d\tau^2+A^2(\tau)\left(1+2\Delta_a-\frac{\bk}{2}R^2\right)\left(dR^2+R^2d\varOmega^2\right),
\end{equation}
where $A(\tau)$ is the scale factor of the FLRW universe model with $\bk=0$ and 
\begin{equation}
\Delta_a:=\frac{a(\tau)-A(\tau)}{A(\tau)} \label{Da-def}
\end{equation}
with the assumption $|\Delta_a|\ll1$. 
By the definition, we have
\begin{equation}
\left(\frac{1}{a}\frac{da}{d\tau}\right)^2=\frac{8\pi\brho_0}{3a^3}-\frac{\bk}{a^2}+\frac{\Lambda}{3}. 
\label{F-eq}
\end{equation}
On the other hand, from Eq.~(\ref{Da-def}), we have
\begin{equation}
\left(\frac{1}{a}\frac{da}{d\tau}\right)^2=\left(\frac{1}{A}\frac{dA}{d\tau}\right)^2
+\frac{2}{A}\frac{dA}{d\tau}\frac{d\Delta_a}{d\tau}.
\label{P-eq}
\end{equation}
We may assume that $A$ satisfies
\begin{equation}
\left(\frac{1}{A}\frac{dA}{d\tau}\right)^2=\frac{8\pi\brho_0}{3 A^3}+\frac{\Lambda}{3}.
\end{equation}
Then, we have, from Eqs.~(\ref{F-eq}) and (\ref{P-eq}),
\begin{equation}
\frac{8\pi\brho_0}{A^3}\Delta_a+\frac{\bk}{A^2}
+\frac{2}{A}\frac{dA}{d\tau}\frac{d\Delta_a}{d\tau}=0.
\label{Dela-eq}
\end{equation}
Here it should be noted that Eq.~(\ref{Dpm-sol}) leads to
\begin{equation}
\left(8\pi \bar{\rho}-\frac{2\bar{k}}{a^2}\right)(aD_+)-\frac{2}{a^2}+2\bar{H}\frac{d (aD_+)}{dt}=0.
\label{Dp-eq}
\end{equation}
By comparing Eq.~(\ref{Dela-eq}) with Eq.~(\ref{Dp-eq}),  we find  
\begin{equation}
\Delta_a=-\frac{\bk}{2}A(\tau){\cal D}_{+}(\tau),
\end{equation}
where
$$
{\cal D}_{+}=\frac{{\cal H}(A)}{A}\int^A\frac{dB}{B^3{\cal H}^3(B)}
$$
with
$$
{\cal H}(B)=\sqrt{\frac{8\pi\brho_0}{3 B^3}+\frac{\Lambda}{3}}.
$$
The density contrast of the FLRW with $\bk\neq0$ relative to that of $\bk=0$ is defined as
\begin{equation}
\Delta_{\rm FLRW}:=\left(\frac{\brho_0}{A^3}\right)^{-1}\left(\frac{\brho_0}{a^3}-\frac{\brho_0}{A^3}\right)
=-3\Delta_a=\frac{3}{2}\bk A(\tau){\cal D}_+(\tau).
\end{equation}
 
Note that $A$ is not unity at present time $\tau=\tau_0$, if $a$ is. 
By the definition, we have 
$$
A(\tau_0)=1-\Delta_{a0},
$$
where $\Delta_{a0}:=\Delta_a(\tau_0)$. We introduce the following quantities; 
\begin{equation}
\ha=\left(1+\Delta_{a0}\right)A,~~~~
r=\left(1-\Delta_{a0}\right)R,~~~~\hat{k}=\left(1+2\Delta_{a0}\right)\bk~~~~{\rm and}~~~~
\hrho_0=\left(1+3\Delta_{a0}\right)\brho_0.
\end{equation}
Then, the infinitesimal world interval becomes
\begin{equation}
ds^2=-d\tau^2+\hat{a}^2(\tau)\left(1+2\Delta_a-\frac{\hat{k}}{2}r^2\right)
\left(dr^2+r^2d\varOmega^2\right),
\end{equation}
The density contrast $\Delta_{\rm FLRW}$ is written as
\begin{equation}
\Delta_{\rm FLRW}=\frac{3}{2}\bk A(\tau){\cal D}_+(\tau)=\frac{3}{2}\hat{k}\hat{a}(\tau)\hat{D}_+(\tau),
\end{equation}
where 
$$
\hD_+(\tau)=\frac{\hH(\ha)}{\ha}\int^{\ha}\frac{dB}{B^3\hH^3(B)}
$$
with
$$
\hH(B)=\sqrt{\frac{8\pi\hrho_0}{3 B^3}+\frac{\Lambda}{3}}.
$$
By the result obtained here, we can 
derive the density contrast $\Delta_{\rm f}$ relative to the FLRW universe model with $\bk=0$ 
from the density contrast $\Delta$ relative to that with $\bk\neq0$ in the manner
\begin{equation}
\fDelta=\Delta+\Delta_{\rm FLRW}=\Delta+\frac{3}{2}\hat{k}\hat{a}(\tau)\hD_+(\tau).
\end{equation}

By using the result obtained in this appendix, 
the infinitesimal world interval (\ref{SC-gauge}) may be rewritten in the form 
\begin{eqnarray}
ds^2&=&-d\tau^2+\ha^2(\tau)\left[1-2(\hpsi+\hat{f})\right]\cr
&& \cr
&\times&
\left[
\left(1-\frac{\ha}{2\pi\hrho_0}\partial_r^2\hpsi\right)dr^2
+r^2\left(1-\frac{\ha}{2\pi\hrho_0}\frac{1}{r}\partial_{r}\hpsi\right)
d\varOmega^2
\right],
\end{eqnarray}
where
\begin{eqnarray}
\hpsi&=&\psi+\frac{2\pi\hrho_0 r^2\Delta_{\rm FLRW}}{3\ha}, \\
\hat{f}&=&f-\frac{2\pi\hrho_0 r^2\Delta_{\rm FLRW}}{3\ha}+\frac{r^2\Delta_{\rm FLRW}}{6\ha\hD_+}.
\end{eqnarray}


\end{document}